\documentclass[twocolumn,showpacs,amsfonts,aps,prc,floatfix,nofootinbib,superscriptaddress,groupedaddress]{revtex4-1}
\usepackage{float}  
\usepackage{array}
\usepackage{booktabs}
\usepackage[colorlinks,linkcolor=red,citecolor=blue]{hyperref}
\usepackage{amsmath}
\usepackage{mathrsfs}
\usepackage{amssymb}
\usepackage{graphicx,epsfig,latexsym,overpic,amssymb,color}
\usepackage{multirow}
\usepackage[section]{placeins}
\hypersetup{
	colorlinks=true,
	urlcolor=blue,
}
\preprint{\today}
\begin{document}
\title{Probing the two-quasiparticle $K^\pi=8^+$ isomeric structure and enhanced stability in the proton drip-line nuclei}

\author{Zhen-Zhen Zhang}
\affiliation{School of Physics,
 Zhengzhou University, Zhengzhou 450001, China}
\author{Hua-Lei
Wang}\email{wanghualei@zzu.edu.cn} \affiliation{School of Physics,
 Zhengzhou University, Zhengzhou 450001, China}
\author{Kui Xiao}
\affiliation{School of Physics,
 Zhengzhou University, Zhengzhou 450001, China}
\author{Min-Liang Liu}
\affiliation{Key Laboratory of High Precision Nuclear Spectroscopy,
Institute of Modern Physics, Chinese Academy of Sciences, Lanzhou
730000, China} \affiliation{School of Nuclear Science and
Technology, University of Chinese Academy of Sciences, Beijing
100049, China}

\begin{abstract}
Stimulated by recent experimental discoveries [\href{https://doi.org/10.1016/j.physletb.2023.138310} {Phys. Lett. B \textbf{847}, 138310 (2023)} and \href{https://doi.org/10.1103/PhysRevLett.132.072502} {Phys. Rev. Lett. \textbf{132}, 072502 (2024)}], two-quasiparticle $K^\pi=8^+$ isomeric structure (related to the neutron $h_{9/2}$ and $f_{7/2}$ orbitals) in $^{160}_{76}$Os$_{84}$ that lies at the two-proton drip line has been studied by means of the configuration-constrained potential-energy-surface calculations. Calculated results indicate that, for such an isomer, the excitation energy can be well reproduced and its oblate shape can be enhanced by the polarization effects of the two high-$K$ orbits. Comparing with experimental data, two sets of the widely used Woods-Saxon parameters, especially, the spin-orbit coupling one, are evaluated and argued.  It is found that, considering the uncertainty of the spin-orbit coupling strength, the energy crossing or inversion of the $h_{9/2}$ and $f_{7/2}$ neutrons can occur, which may lead to three kinds of different evolution-trends of two-quasiparticle excitation energies with the changing quadrupole deformation $\beta_2$. With decreasing spin-orbit coupling interaction, the structure of the $K^\pi=8^+$ isomeric state will evolute from $\nu h_{9/2}f_{7/2}$ ($\nu 9/2^-[505] \otimes 7/2^-[503]$) to the mixing of $\nu h_{9/2}f_{7/2}$ and $\nu h_{9/2}^2$ ($\nu 9/2^-[505] \otimes 7/2^-[514]$)  to $\nu h_{9/2}^2$, indicating that its structural probes is still of interest and an arbitrary assignment may be risky. The related theoretical calculations and experimental evidences e.g., the transition properties, are desirable. In addition, similar to that in superheavy nuclei, it is suggested that the stability inversion between high-$K$ isomeric states and ground states might occur in this proton drip-line mass region, e.g., in the hitherto unknown nucleus $^{162}_{78}$Pt$_{84}$.

\end{abstract}
\maketitle

\section{INTRODUCTION}
\label{Introduction}
	
The atomic nucleus is a quantum many-body system comprising two types of nucleons, the neutron and proton fermions. The different arrangement of these nucleons within the same nucleus, involving different orbital configurations, may produce nuclear isomers who are metastable nuclear excited states with long half-lives, e.g., greater than one nanosecond (in general, nuclear excited states with typical picosecond or femtosecond lifetimes)~\cite{Walker2024}. For instance, the hafnium nucleus $^{178}$Hf contains 178 nucleons (72 protons and 106 neutrons) but has zero angular momentum (due to the pairing interactions) at its stable ground state. However, when just two protons and two neutrons rearrange, a four-quasiparticle $K^\pi = 16^+$ isomer forms, which has the 31-year half life and 2.4-MeV excitation energy. In fact, such nuclear isomers are usually caused due to the hindrance of their decay processes back to the ground state or lower excited-state(s). Moreover, the decay process will depend more-or-less entirely on spontaneous emissions since the nucleus, as an almost isolated object, cannot de-excite rapidly by nucleus-nucleus collisions, e.g., like the chemical isomer de-excitations in atoms~\cite{Walker1999}. Based on the different physical constraints, the nuclear isomers can be classified into spin traps~\cite{Xu2000,Dracoulis2006}, $K$ traps~\cite{Jeppesen2009,Xu2004,Liu2011,liu2015} and shape isomers ~\cite{Alkhomashi2009,Walker2006,Guan2025}. The large change of angular momentum, usually accompanying low transition energy, will lead to the occurrence of extremely low electromagnetic transition rates in the formers. The large change in the direction of the angular momentum and the significant shape change of the nucleus are respectively responsible for $K$ and shape isomers, where the excited-state's decay is also strongly inhibited. These isomers are crucial for probing, e.g., nuclear structure, nuclear astrophysics, even new energy-storage forms and $\gamma$-ray lasers~\cite{Misch2021,Walker1999}. Furthermore, they could be a bridge between nuclear and atomic physics due to the nucleus-electron interactions, including the effects of internal conversion (IC), nuclear excitation by free electron capture (NEEC) and nuclear excitation by electron transition (NEET)~\cite{Walker2022,Cai2021,Si2025,Yang2025,Liu2025}. 

Since the first isomer was discovered in $^{234}$Pa by Hahn~\cite{Hahn1921}, thanks to advances in experimental techniques, including nuclear detectors, rare beam facilities, reaction analyzers, digital electronics, and so on, numerous isomers, even their detailed properties, have been observed. For instance, Jain et al., listed almost 2500 isomers with half-lives greater than 10 ns~\cite{Jain2015}. In the recent NUBASE2020 evaluation~\cite{Kondev2021}, 1938 isomers with half-lives greater than 100 ns were collected. In the present project, our primary concern is the multi-quasiparticle high-$K$ isomers whose decay to low-$K$ states is inhibited due to $K$ forbiddenness~\cite{Walker1999}, where $K$ is the total angular momentum projection onto the nuclear symmetry axis.
Such multi-quasiparticle isomeric states can provide essential information about the nucleonic orbits close to the Fermi surfaces, which may play an important role in testing nuclear structure models (including model parameters) and further developing them. 

Indeed, both high-$K$ isomers and drip-line nuclei, especially their combinations, attract extensive research interest in nuclear physics, e.g., owing to continuous development of radioactive ion beam facilities~\cite{Thoennessen2011}.  Presently, instead of providing a review of nuclear isomers (but the interested reader can consult review articles, e.g., see Refs.~\cite{Walker1999, Walker2001, Dracoulis2016} and references therein), we will focus on probing the two-quasiparticle high-$K$ isomeric structure in the two-proton drip-line nucleus $^{160}_{76}$Os$_{84}$ which was reported in two recent papers~\cite{Briscoe2023,Yang2024}, where the new nuclide  $^{160}_{76}$Os$_{84}$  was produced using the same fusion-evaporation reaction $^{106}$Cd($^{58}$Ni,4$n$)$^{160}$Os but different beam energy (310 and 335 MeV, respectively). In these two parallel publications, the ground-state $\alpha$-decay energies of $^{160}$Os are respectively 7095(15) and 7080(26) keV, agreeing with each other very well and the measured half-lives of the ground states ($t_{1/2} = 97^{+97}_{-32}$ and $210^{+58}_{-37}$ $\mu s$) are also in agreement within the error bars. Nevertheless,  the 8$^+$ isomeric state with the excitation energy 1.844 (18) MeV and the half life about $41^{+15}_{-9}$ $\mu s$ was identified in Ref.\cite{Briscoe2023} but no evidence was found for the isomer in Ref.~\cite{Yang2024}. Information concerning the excited-state structure is relatively scarce in these nuclei far from the $\beta$-stability line. The testing of structure calculations has also been limited and further theoretical investigation in these drip-line nuclei may be of interest and necessary.

Prior to this work, following theoretical predictions in the $N= 84$ isotones, the systematic existence of two-quasiparticle $K^\pi = 8^+$ state interpreted in terms of neutron excitations involving the $\nu f_{7/2}h_{9/2}$ configuration has been demonstrated in $^{154}_{70}$Yb$_{84}$, $^{156}_{72}$Hf$_{84}$ and $^{158}_{74}$W$_{84}$ by experiments ~\cite{Zhang1993,Joss2017,Seweryniak2005}. Naturally, along the $N=84$ isotonic chain, the $K^\pi = 8^+$ isomer is expected to exist in $^{160}_{76}$Os$_{84}$. However, it was pointed out, e.g., in Ref.~\cite{Zhang1993}, that the lowering of the $\nu h_{9/2}$ single particle state is a characteristic structure feature in the $N=83$ and $N=84$ isotones. Moreover, it was expected that the $\nu 1h_{9/2}$ state would drop below $\nu 2f_{7/2}$ in the $^{159, 160}_{76}$Os$_{83, 84}$. This may affect the isomer structure of the two-quasiparticle $K^\pi =8^+$ state.  As is known, the orbits  $2f_{7/2}$ and $1h_{9/2}$ respectively correspond to the $l+1/2$ and $l-1/2$ signatures of the spin doublets ($n,l,j=l \pm 1/2$). 
The relative positions of such a pair of levels will be strongly affected by the spin-orbit coupling which possesses a rather large uncertainty~\cite{Zhang2021}. One of our aims in this project is to explore the effects of spin-orbit coupling on the intrinsic configuration of the two-quasiparticle $K^\pi=8^+$ isomeric states, especially, in $^{160}_{76}$Os$_{84}$ and $^{162}_{78}$Pt$_{84}$.

In addition, it is well known as a remarkable feature that extra stability might be conferred due to the high angular momentum in some excited nuclear states which can have longer half-lives than their respective ground states~\cite{Walker2012}. For instance, for the tantalum nucleus $^{180}_{73}$Ta, a quasi-stable $K^\pi = 9^-$ isomer has a half-life of more than $10^{15}$ years and the excitation energy of just 75 keV but the ground-state half-life is just 8 hours~\cite{Audi2003}. The $^{256}_{99}$Es $8^+$ isomer has a half-life of 7.6 h, which is much longer than the 25 min of the corresponding $1^+$ ground state~\cite{Xu2004,Audi2003}.
In the heavy nucleus $^{250}_{102}$No, the ground state has the shorter half life ($t_{1/2} = 3.7^{+1.1}_{-0.8} \mu s$) while the longer decay ($t_{1/2} = 43^{+22}_{-15} \mu s$) is attributed to a $K^\pi = 6^+$ isomeric state~\cite{Peterson2006}.
In the superheavy nucleus $^{270}_{110}$Ds, an isomeric state with $K^\pi=9^-$ (or $10^-$) was observed with a half-life of $6.0^{+8.2}_{-2.2} ms$ that is significantly longer than the $100^{+140}_{-40} \mu s$
of the corresponding ground state~\cite{Hofmann2001}. Very recently, the new superheavy nuclide $^{252}_{104}$Rf was identified and its ground-state decay as well as the decay of a high-$K$ isomeric state with half-lives of $60^{+90}_{-30} ns$ and $13^{+4}_{-3} \mu s$ were observed, further indicating the stability inversion between the high-$K$ and the ground states~\cite{Khuyagbaatar2025,Zhang2025}. For some superheavy nuclei with sufficiently short lifetimes, the stability inversion might occur between high-$K$ isomers and ground states; namely, the half lives of the excited states can be longer lived than the corresponding ground states and these effects will favour the synthesis of superheavy nuclei~\cite{Xu2004,Walker2012}. In the drip-line region, it is of interest to probe whether the broken-pair, high-K isomers may confer extra stability. In particular, to look for the possible candidate for which the stability inversion occurs between the high-$K$ isomeric state and the ground state in this nuclear region will be our another research purpose in this project.

At least, keeping these two goals mentioned above in mind, we will investigate the structure properties of the two-quasiparticle $K^\pi=8^+$ isomeric states and the possible stability inversion in the neutron-deficient $N=84$ isotones based on the configuration-constrained potential-energy-surface (PES) calculation in the multidimensional deformation space ($\beta_2, \gamma, \beta_4$)~\cite{Xu1998} and the evalution of their $\alpha$-decay half lives. We are also interested in testing the applications of phenomenological Woods-Saxon mean-field model and model parameters (e.g., the uncertainty of the spin-orbit coupling strengths and its effect on nuclear configurations).    

	
The paper is organized as follows: In Sect. 2, we briefly introduce the theoretical framework of configuration-constrained potential-energy-surface calculation and some possibly helpful points for the general readers.
In Sect. 3, we show the calculation results related to our research motivations, discussing the model parameters, the intrinsic structures of the two-neutron $K = 8^{+}$ isomer and the problem of enhanced stability.
Finally, we provide a brief summary and outlook in Sect. 4 for the present study.

\section{Theoretical method}
	
In this section, we will summarize the leading lines and basic definitions (including some points which may render confusions, e.g., the pairing energy and the blocking way) related to the configuration-constrained PES calculations~\cite{Xu1998} adopted in this project and simultaneously provide some necessary references.

Nuclear potential energy is calculated within the framework of macroscopic-microscopic Strutinsky method with a realistic phenomenological
mean-field Hamiltonian~\cite{Xu1998,Meng2022,Guo2026}. We describe the nuclear shape by means of the parameterization of the nuclear surface with the help of spherical harmonics (Bohr's parameterization is also adopted~\cite{Bohr1988}). A three-dimensional deformation space ($\beta_2, \gamma, \beta_4$) is adopted, including the low-order axial and nonaxial deformation degrees of freedom. Such a deformation space is widely used in the literature~\cite{Xu1998}. The microscopic single-particle levels are obtained from the deformed Woods-Saxon potential and parametrization from Ref.~\citep{Cwiok1987,Bhagwat2010,Bhagwat2021,Bhagwat2023}. To avoid the spurious pairing phase transition encountered in the usual BCS approach, we use the approximate particle-number conserved Lipkin-Nagami (LN) treatment of pairing~\cite{Pradhan1973}, with monopole pairing considered and the pairing strength $G$ determined by the average-gap method~\cite{Moller1992}. The pairing windows for both protons and neutrons contain dozens of single-particle levels, e.g., half of the nucleon number $Z$ and $N$, above and below the Fermi surface.
	
The total energy of a given configuration consists of a macroscopic part, which is obtained from the standard liquid-drop model~\cite{Myers1966}, and a microscopic part resulting from the Strutinsky shell correction~\cite{Strutinsky1967}, $\delta E_{\text{shell}}=E_{\text{LN}}-\tilde{E}_{\text{Strut}}$. Such a microscopic part, as discussed in the literature, e.g., see Ref.~\cite{Gaamouci2021}, is equivalent to the summation of pairing correlation, $E_{\text{LN}}(\Delta \neq 0) - E_{\text{LN}}(\Delta = 0)$, and shell correction, $E_{\text{LN}}(\Delta = 0) - \tilde{E}_{\text{Strut}}$.   Here, the LN energy for a configuration-constrained multi-quasiparticle state of an even-even nucleus at ``paired solution'' (pairing gap $\Delta\neq 0$) is given by~\cite{Pradhan1973,Moller1995}
\begin{eqnarray}
	E_{LN} (\Delta \neq 0)&=&\sum_{j=1}^{S}e_{k_{j}}+\sum_{k\neq k_{j}}2{v_k}^2e_k
	-\frac{{\Delta}^2}G-G\sum_{k\neq k_{j}}{v_k}^4
	\nonumber \\
	&& +G\frac{N-S}{2}-4{\lambda}_2\sum_{k\neq k_{j}}({u_k}{v_k})^2,
	                                                                \label{eqn.01}
\end{eqnarray}
where $S$, always an even number, is the proton or neutron seniority (namely, the unpaired particles, i.e. the number of blocked orbitals with index $k_{j}$) and $N$ is the proton or neutron number. Correspondingly, the
partner expression at ''no-pairing solution'' ($\Delta = 0$) reads
\begin{eqnarray}
	E_{LN}(\Delta=0)&=&\sum_{j=1}^{S}e_{k_{j}}+\sum_{k\neq k_{j}}2e_k.
                      	                                            \label{eqn.02}
\end{eqnarray}
The unpaired nucleon orbitals with index $k_j$ are used for specifying the constrained  multi-quasiparticle configuration. Note that the index $k$ only involves the occupied orbitals at this moment. At each point deformation lattice of the selected multidimensional deformation space, these occupied orbitals involved in the given configuration will be always tracked, e.g., by calculating and identifying the average Nilsson quantum numbers, and blocked to participate the LN-pairing calculation (that is, the so-called configuration constraint)~\cite{Xu1998}. 

It should be noted that there are two ways, adiabatic and diabatic ones, to block the unpaired nucleon orbitals during the configuration-constrained PES calculations. The adiabatic case does not consider the spurious avoided crossings of calculated single-particle levels, while the later considers exchange of the wave functions at the points of avoided crossings and removal of the virtual interactions, breaking the noncrossing rule -- the eigenvalues of eigenstates of the same symmetry will not cross, e.g., cf. Refs.~\cite{Hatton1976,Bengtsson1988} and references therein. The spurious avoided crossing, the invalidity of adiabatic approximation as solving the Schrödinger equation at this moment, generally origintes from the missing symmetries of the model Hamiltonian and the limitation of the used basis space~\cite{Hatton1976}. Of course, there is no difference for such two blocking ways in the region far from avoided-crossing points. At present, the level-identification way by the average Nilsson quantum numbers can usually track the diabatic orbitals very well, especially in a small deformation domain~\cite{Xu1998}.      
	
Similarly, as described in Ref.~\cite{Meng2022,Xu1998}, one can obtain the  configuration-constrained potential energy at each sampling deformation grid. Further, the smooth  PES/map can be given with the help of the interpolation techniques, e.g., a cubic spline function and then the equilibrium
deformations and other physical quantities, e.g., excitation energy and pairing property of a constrained multi-quasiparticle state at such a shape, can be determined. Let's stress that the excitation energy of multi-quasiparticle configuration, which is calculated by the energy difference between the minima of the configuration-constrained and ground-state PESs, can be decomposed into the deformation energy and the configuration energy that corresponds to the quasiparticle excitations that undergo the pairing breaking and the particle excitaitons~\cite{Shi2010}. Once the large shape change occurs due to the polarization effect of the unpaired nucleons, the deformation energy will not be ignored. Indeed, the configuration-constrained PES calculation, self-consistently treating the deformation, pairing and energy, has become a powerful tool in the study of multi-quasiparticle high-$K$ isomeric states~\cite{Shi2010, Xu1998}.

\section{Results and Discussion}
\subsection{Evaluation and selection of model parameters}

There exist various parameterizations of the WS potential, e.g., the Blomqvist, Chepurnov, Rost, universal and Stockholm parameter sets~\cite{Wang2015,Cwiok1987,Bhagwat2023}, but no one is perfect. Moreover, the WS parametrizations never stop developing both in nuclear structure and reaction, e.g., see Refs.~\cite{Gaamouci2021,Gan2017}. One optimal parameter set for the global region of the nuclear chart, e.g., the universal~\footnote{As given in Ref.~\cite{Yang2022}, the adjective ``universal'' means the fact that the underlying parameterization is applied to all the nuclei of the nuclear chart without further modifications.} parameterization~\cite{Yang2022}, may be not the best one for a local nuclear region.  

\begin{table}
	\caption{
		\label{table1}
	Calculated Euclidean ($d_E$), Manhattan ($d_M$) and Chebyshev ($d_C$) distances distances between $^{160}_{76}$Os$_{84}$ and 8 doubly magic spherical nuclei; see the text for more details.
	}
\begin{ruledtabular}
\begin{tabular}{lccc}
\specialrule{0em}{2pt}{2pt}
Nuclei &  $d_E$ & $d_M$ & $d_C$  \\	
\specialrule{0.04em}{2pt}{2pt}
$^{16}_{8}$O$_{8}$   &102.0 & 144 & 76 \\
\specialrule{0em}{2pt}{2pt}
$^{40}_{20}$Ca$_{20}$  &85.0 & 120 & 64 \\
\specialrule{0em}{2pt}{2pt}
$^{48}_{20}$Ca$_{28}$  &79.2 & 112 & 56 \\
\specialrule{0em}{2pt}{2pt}
$^{56}_{28}$Ni$_{28}$  &73.8 & 104 & 56 \\ 
\specialrule{0em}{2pt}{2pt}
$^{90}_{40}$Zr$_{50}$  &49.5 & 70 & 36 \\ 
\specialrule{0em}{2pt}{2pt}
$^{132}_{50}$Sn$_{82}$ &26.1 & 28 & 26  \\   
\specialrule{0em}{2pt}{2pt}
$^{146}_{64}$Gd$_{82}$ &12.2 & 14 & 12 \\  
\specialrule{0em}{2pt}{2pt}
$^{208}_{82}$Pb$_{126}$&42.4 & 48 & 42 \\ 
\specialrule{0em}{2pt}{2pt}
\end{tabular}
\end{ruledtabular}
\end{table}

In general, the WS mean-field parameters are fitted to the experimental values of single-particle energies in part or all of the eight spherical doubly-magic nuclei
\begin{eqnarray}
 ^{16}\rm{O}, ^{40}\rm{Ca}, ^{48}\rm{Ca}, ^{56}\rm{Ni}, ^{90}\rm{Zr}, ^{132}\rm{Sn}, ^{146}\rm{Gd}, ^{208}\rm{Pb};	\nonumber
\end{eqnarray}
the corresponding energies for protons and neutrons were extracted in Ref.~\cite{Dudek2010},  
where the empirical values of the single-nucleon energies could be determined from their single-nucleon separation energies $S(n)$ or $S(p)$ in odd-$A$ nuclei. 
Adjacent nuclei often exhibit commonalities in their structural characteristics and properties (e.g. the arrangement patterns of nucleons and shell structures). The distance between two points, representing two nuclei in the nuclear chart, captures the similarity between them. In mathematics and machine learning, as is known, the Minkowski distance is a generalized metric that unifies various distance measures in an $n$-dimensional space. Between two points $P = $ ($x_1, x_2, \cdots, x_n$) and $P^\prime = $ ($x^\prime_1, x^\prime_2, \cdots, x^\prime_n$), such a distance is defined as follow 
\begin{equation}
d(P,P^\prime)=\sqrt[k]{\sum_{i=1}^n |x_i-x^\prime_i|^k}.
	                                                             \label{eqn.03}
\end{equation}
If the parameter $k$ equals to 1, 2 or $+\infty$, the Minkowski distance will be equivalent to the Manhattan, Euclidean or Chebyshev distance, respectively. These three ones are commonly used distance metrics which are useful in various use cases and differ in some important aspects.

To measure the similarities between our target nucleus ($Z_t$, $N_t$), e.g., $^{160}$Os, and each spherical doubly-magic nucleus ($Z_m$,$N_m$) on the chart of nuclides, the calculated Euclidean, Manhattan and Chebyshev distances are illustrated in Table~\ref{table1}. For instance, the Euclidean distance between $^{160}_{76}$Os$_{84}$ and $^{208}_{82}$Pb$_{126}$ [namely, two points $P(76, 84)$ and $P^\prime (82, 126)$ on the nuclide chart] can be calculated by $d_{E} = \sqrt{(76-82)^2+(84-126)^2} \approx 42.4$. As one can see, all of the distances indicate that the closest neighbour of the nucleus $^{160}$Os is $^{146}$Gd in these doubly magic nuclei though no one can be sure which distance is absolutely valid for the present similarity measurement. It may be reasonable to believe that the WS parameter set, which can describe the nucleus $^{146}$Gd better, is also an optimal candidate for $^{160}$Os. From Table~\ref{table1}, it can be noticed that the ranking is not always same for these three kinds of distances. For instance, the third closest neighbour of $^{160}$Os is the $^{90}$Zr nucleus based on the Chebyshev distance; while it is $^{208}$Pb if using Euclidean and Manhattan distances.

\begin{figure}[H]
\centering
\includegraphics[height=0.33\textwidth]{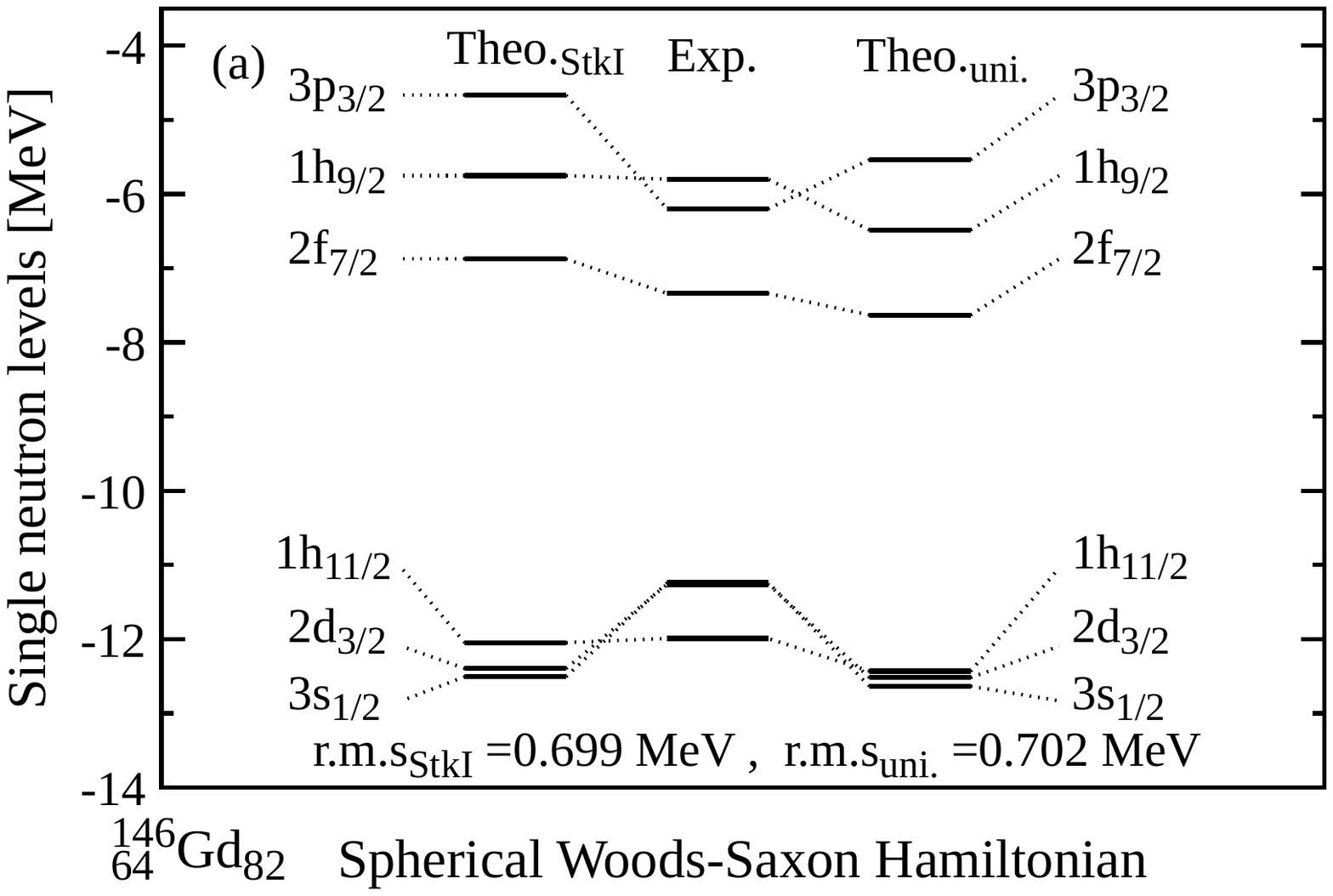}\\
\vspace{-0.3cm}
\includegraphics[height=0.33\textwidth]{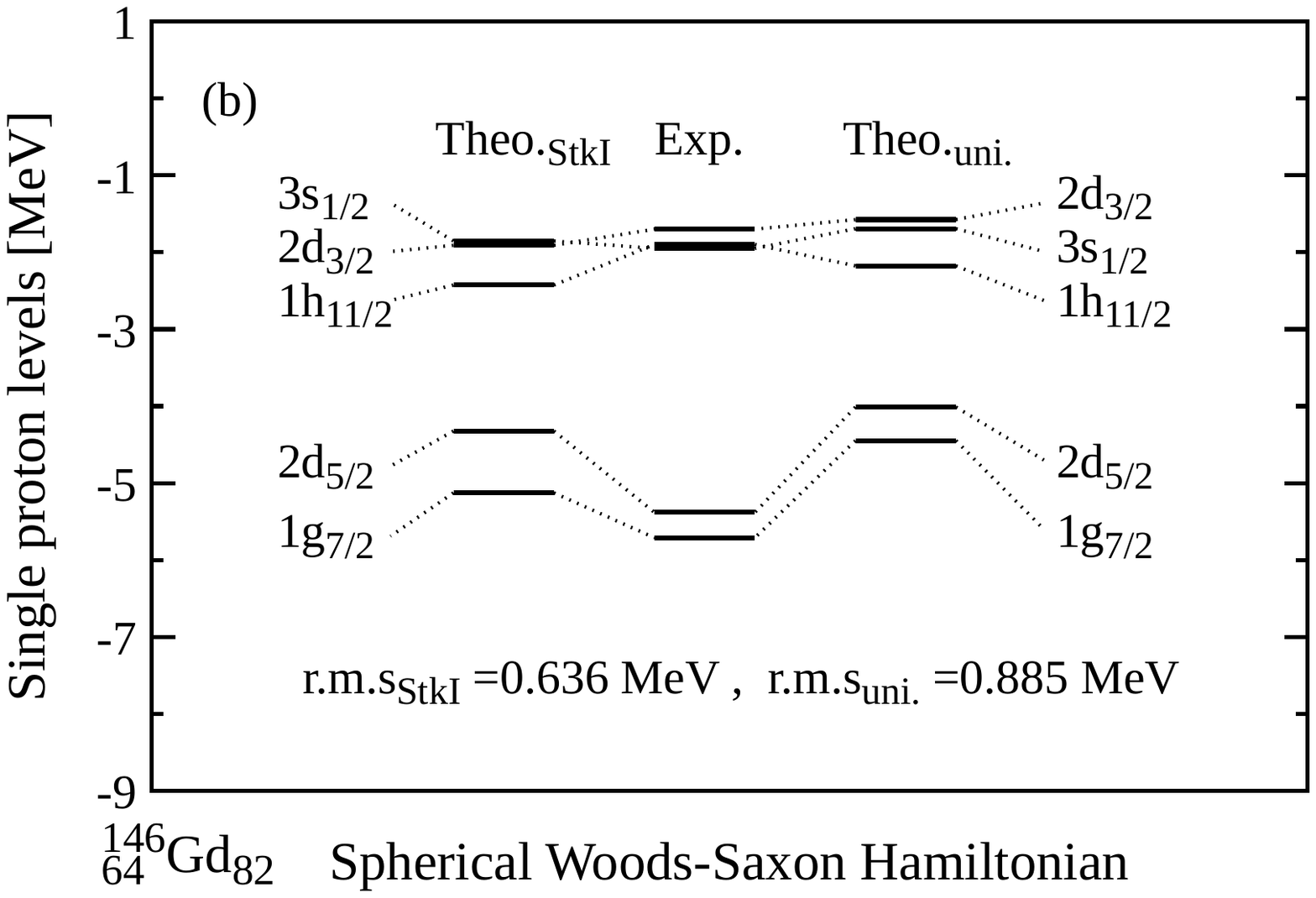}
\vspace{-0.2cm}
\caption{
Comparison between the available experimental data and the calculated single-particle energies in $^{146}$Gd. Note that theoretical single-particle levels are solved by using the Woods-Saxon Hamiltonian with uni. and StkI model parameters, .
		                                                          \label{fig1}
	}
\end{figure}

Note that our primary concern in this project is to probe the high-$K$ isomeric structure rather than to provide a new WS parameter set. Therefore, based on two sets of widely-used WS parameters, namely, universal (uni.) and Stockholm-I (StkI) ones~\cite{Cwiok1987,Bhagwat2010,Bhagwat2023}, we try to evaluate their performance in such a local nuclear region, e.g., the $A\approx 160$ mass region. To a large extent, the model parameters, who describe the experimental observations better, will be more suitable. As shown in Fig.~\ref{fig1}, the comparison between the available experimental single-particle levels and the corresponding theoretical values in $^{146}$Gd is presented, together with the root-mean-square (r.m.s) deviations (one of the important metrics to measure the goodness of the fit). It can be clearly seen that, comparing with experimental data, the StkI parameter set exhibits the relatively good performance, with the smaller r.m.s. deviations, for both protons and neutrons.
It is worth pointing out that the high-$j$ single-particle level calculated using the StkI parameter set, e.g.$\nu h_{9/2}$, is in good agreement with experiment, as seen in Fig.~\ref{fig1}(a). This single-particle orbit is critical for the present study of two-quasi-neutron 8$^+$ isomeric state. 

\begin{figure}[htbp]
\centering
\includegraphics[width=0.45\textwidth]{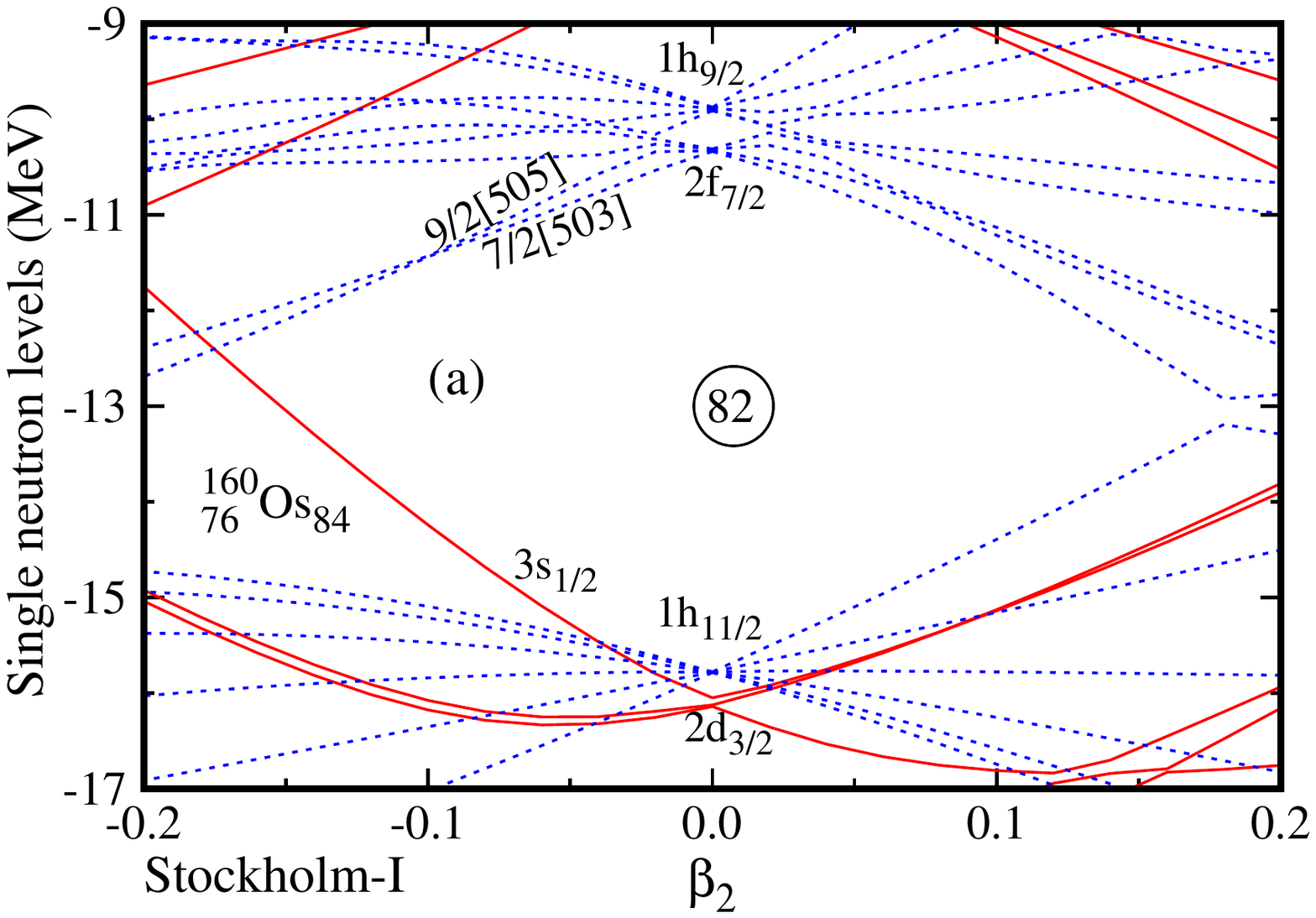}
\includegraphics[width=0.45\textwidth]{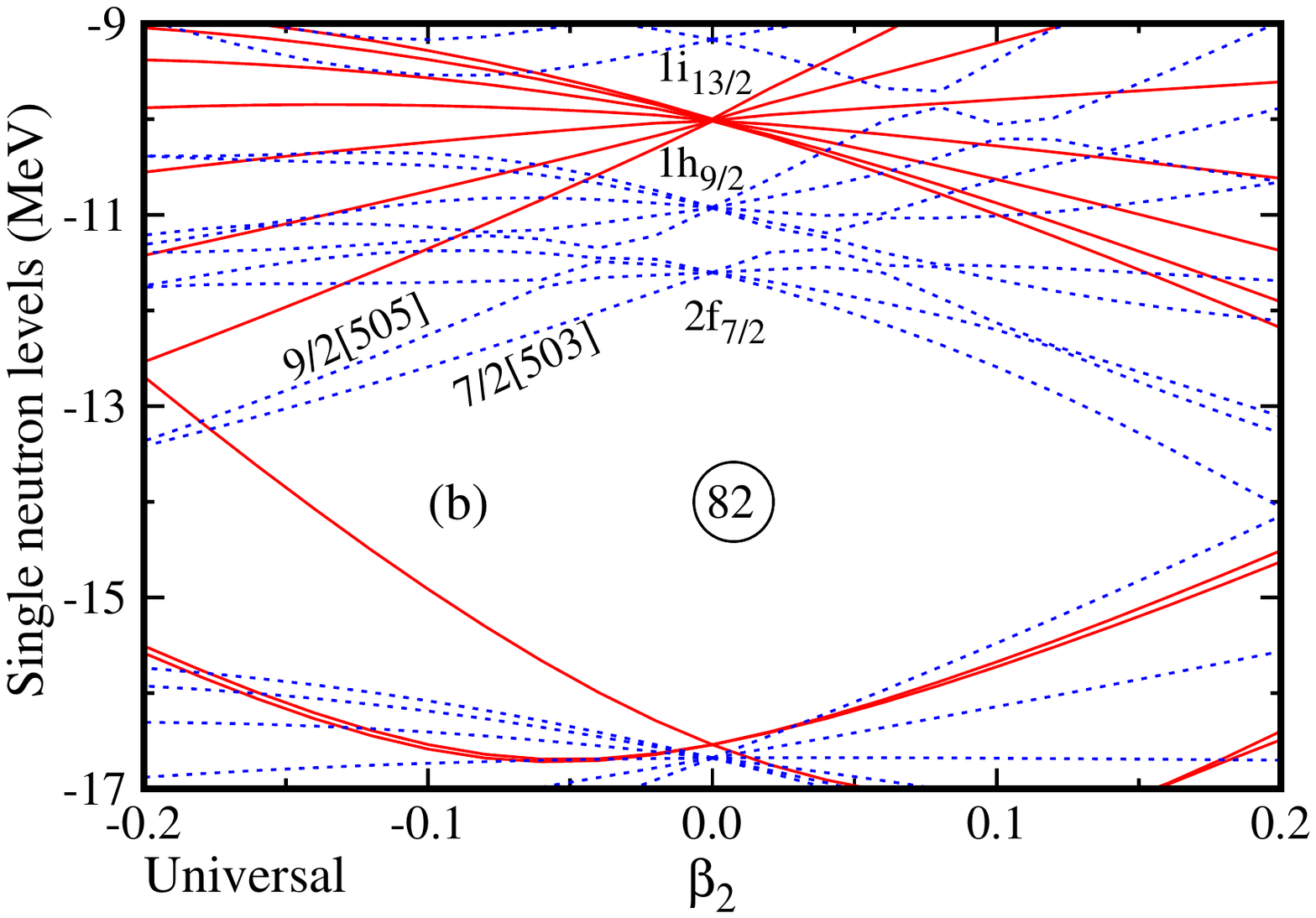}
\caption{ 
Calculated single-particle energies near the Fermi surface for neutrons (similarly for protons) as functions of quadrupole deformation $\beta_{2}$ by using the Woods-Saxon Hamiltonian with the StkI (a) and universal (b) parameters~\cite{Meng2018,Yang2016,Bhagwat2010,Bhagwat2023} for $^{160}$Os. The energy levels with positive and negative parities are respectively denoted by red solid and blue dotted lines. At $\beta_2 = 0.0$, the spherical quantum numbers $nlj$ are adopted for the single-particle labels. In the present deformation space, the single-particle levels at the positive and negative $\beta_2$ values are calculated by setting the triaxial deformation $\gamma= 0^\circ$ and $+60^\circ$ (or $-60^\circ$), respectively. See text for more details.
		                                                         \label{fig2}
	}
\end{figure}

Indeed, the neutron single-particle levels, especially near the Fermi level, have an important influence on the nuclear microscopic properties~\cite{Baldo2020}, e.g., the isomeric structure of two-quasi-neuton configuration. 
Figure 2 illustrates such neutron single-particle levels for $^{160}_{76}\rm{Os}_{84}$ by using the deformed Woods-Saxon potential with two sets of WS parameters, similar to Fig.~\ref{fig1}. Associated with a complete set of commuting observables, a set of conserved quantum numbers is usually used for labeling the corresponding single-particle levels. In the spherical case, for instance, the single-particle levels are denoted by the quantum numbers $n, l$ and $j$, corresponding to the principal quantum number, the orbital angular momentum, and the total angular momentum, respectively.
For the deformed single-particle levels, e.g., at $\beta_2 \neq 0$, the asymptotic Nilsson quantum numbers $\Omega[Nn_{z}\Lambda]$ are usually adopted, where $N$, $n_{z}$, $\Lambda$ and $\Omega$ respectively stand for the total oscillator shell quantum number, the number of oscillator quanta in the $z$ direction, the projection of orbital angular momentum along the symmetry axis and the projection of total angular momentum. In this figure, two interested neutron orbitals $\nu\frac{7}{2}[503]$
and $\nu \frac{9}{2}[505]$ are demonstrated. It can be seen from Fig.~\ref{fig2}(a) that the level crossing of the $\nu\frac{7}{2}[503]$
and $\nu \frac{9}{2}[505]$ orbitals occurs at weakly oblate deformation, e.g., at $\beta_{2} \approx -0.1$, while it is not the case in Fig.~\ref{fig2}(b). The relative positions of these two single-particle levels may be quite different along $\beta_2$. Taking the report in Ref.~[21] into account, we prefer to select the StkI parameter set for the following calculations in this work since it was pointed out that, relative to the $\nu f_{7/2}$ level, the energy of $\nu h_{9/2}$ level has a decreasing trend (namely, the energy difference between $\nu h_{9/2}$ and $\nu f_{7/2}$ orbitals decreases) in the $N=84$ isotones from $Z=64$ to $76$, even smaller than that of $\nu f_{7/2}$, approaching the present case.
%

\begin{figure}[htbp]
\centering
\includegraphics[height=0.33\textwidth]{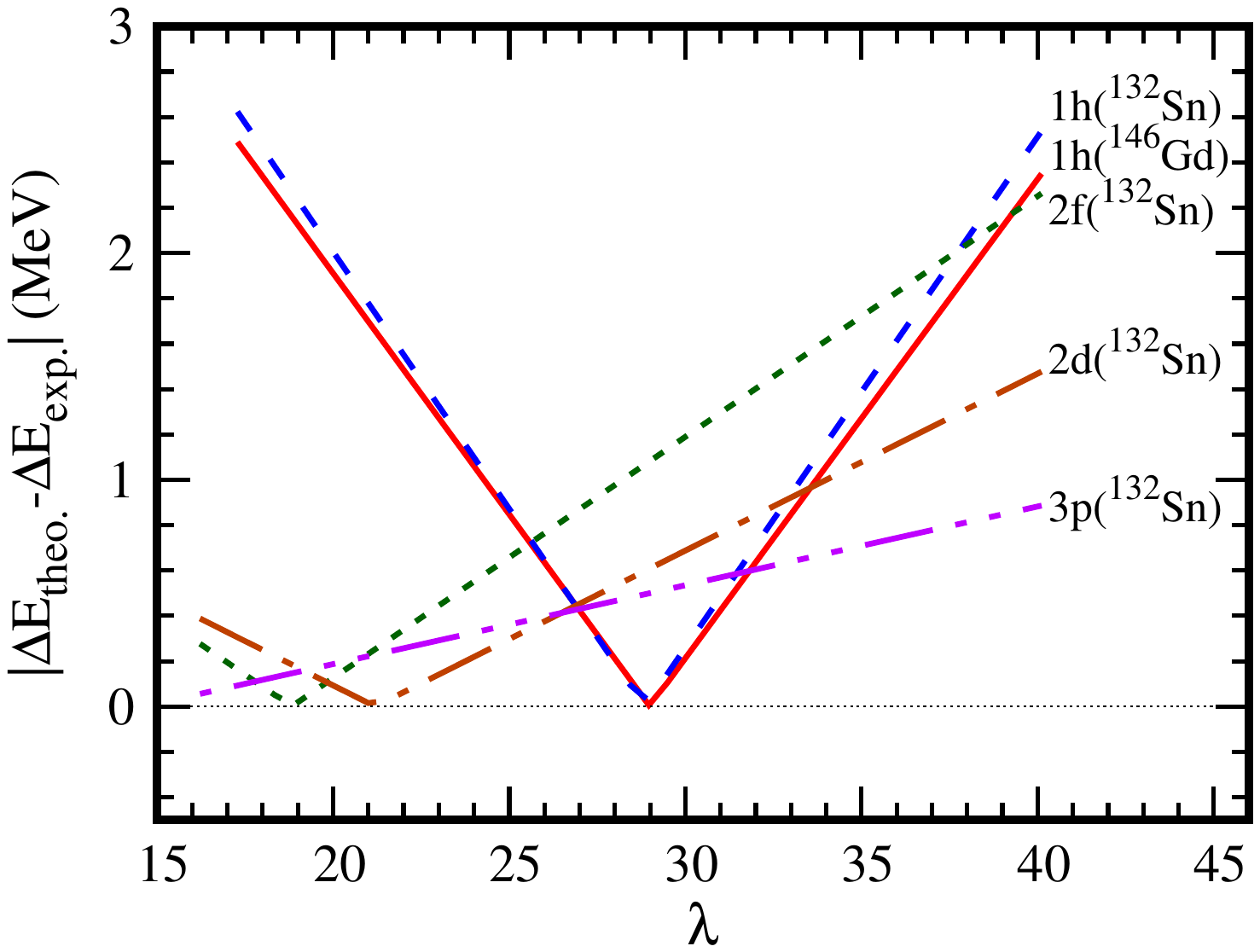}
\caption{ 
Discrepancies between experimental and calculated $l$-level (labeled as \{nl\} quantum numbers) splittings, due to the spin-orbit interaction, as a function of spin-orbit coupling strength $\lambda$ for $^{132}$Sn and $^{146}$Gd. Theoretically calculated values $\Delta$E$_{theo.}$ are obtained based on the Woods-Saxon Hamiltonian with the StkI parameter set. Note that one can see, e.g., Eq.(2.3.1) in Ref.[34] for the definition of spin-orbit potential and the $\lambda$ value is 29.494 in the StkI parameter set. More information see text.
	                                                                 \label{fig3}
	}
\end{figure}

\begin{figure}[htbp]
	\centering
	\includegraphics[width=0.23\textwidth]{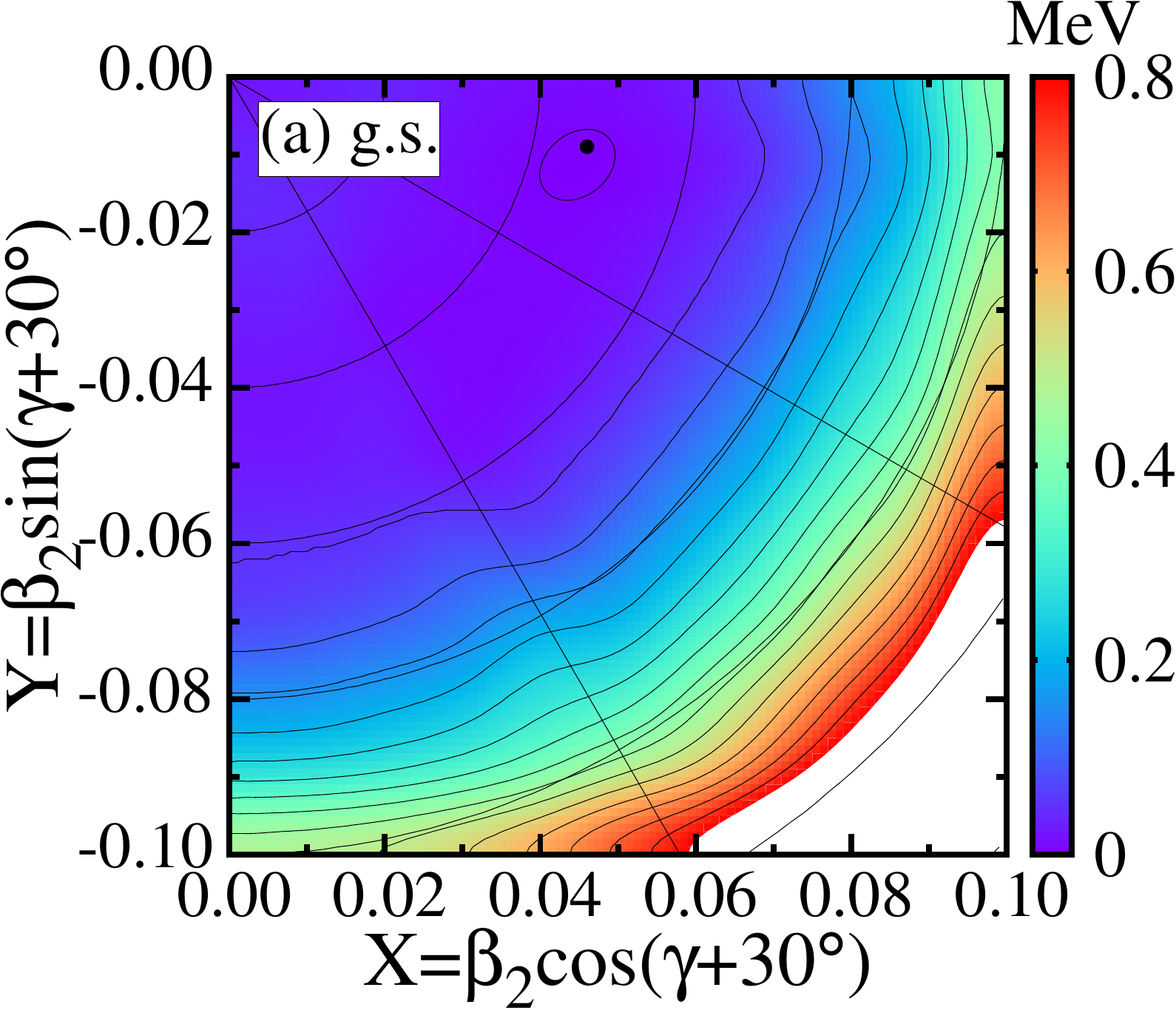}
	\includegraphics[width=0.23\textwidth]{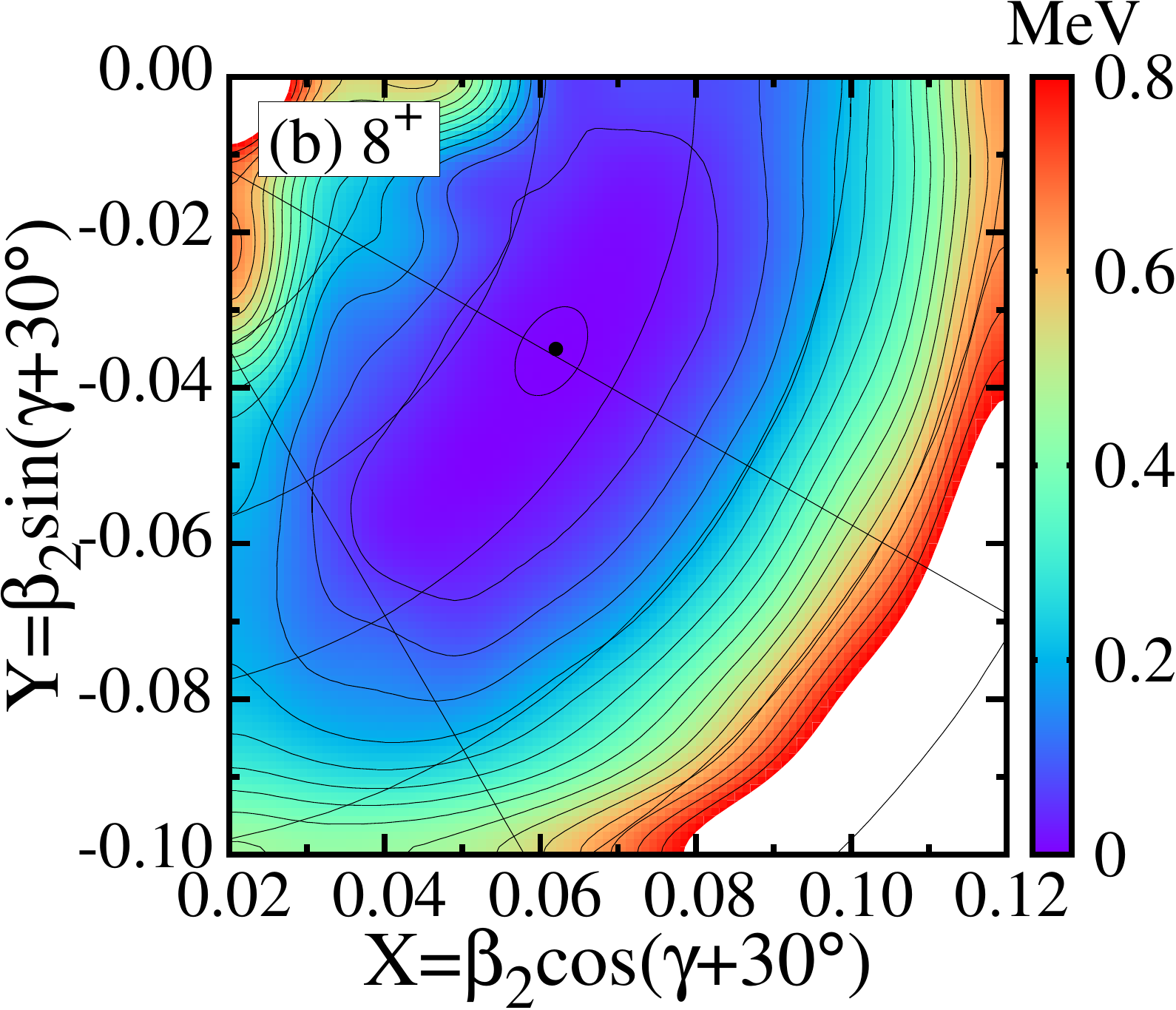}\\
	\includegraphics[width=0.23\textwidth]{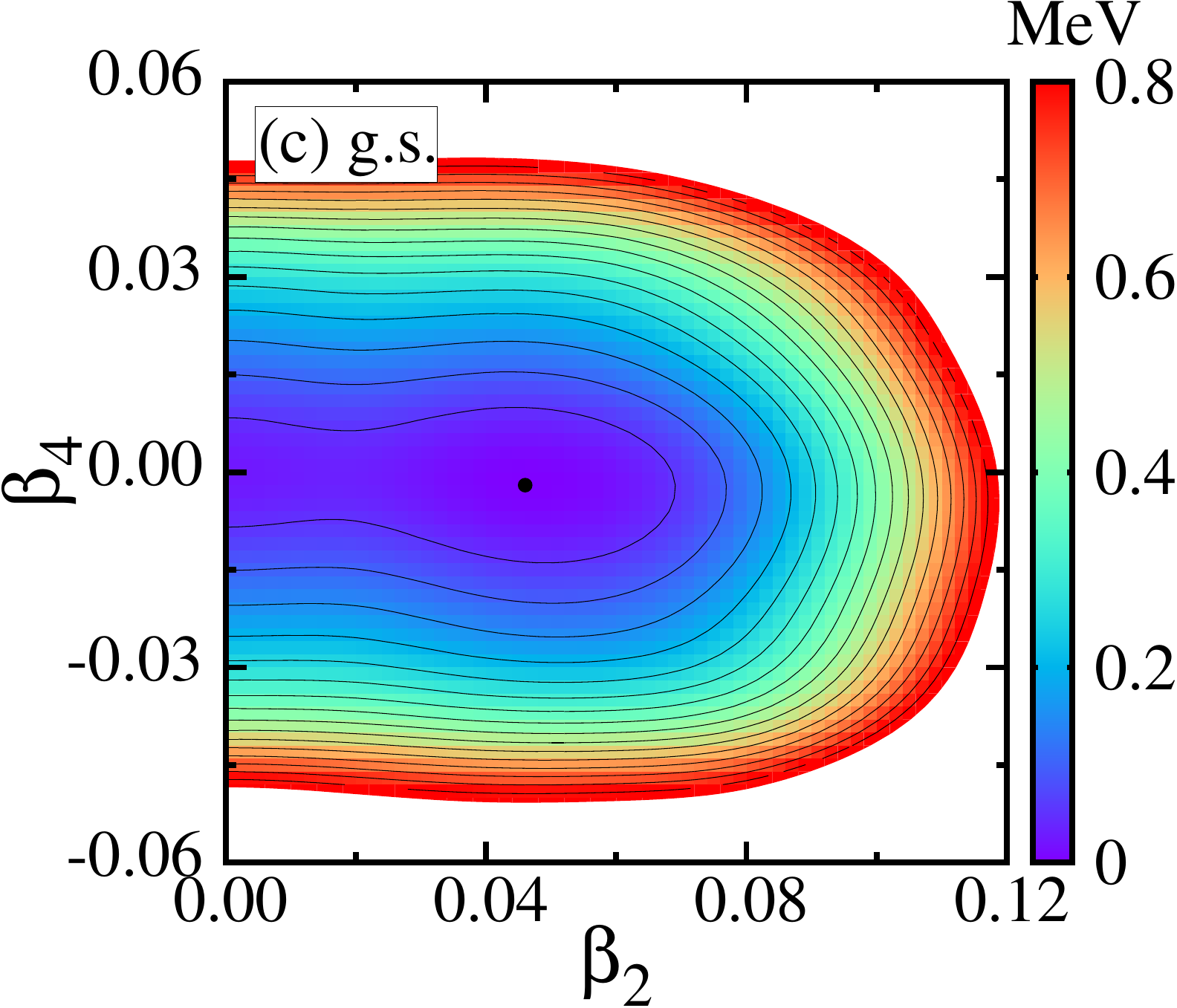}
	\includegraphics[width=0.23\textwidth]{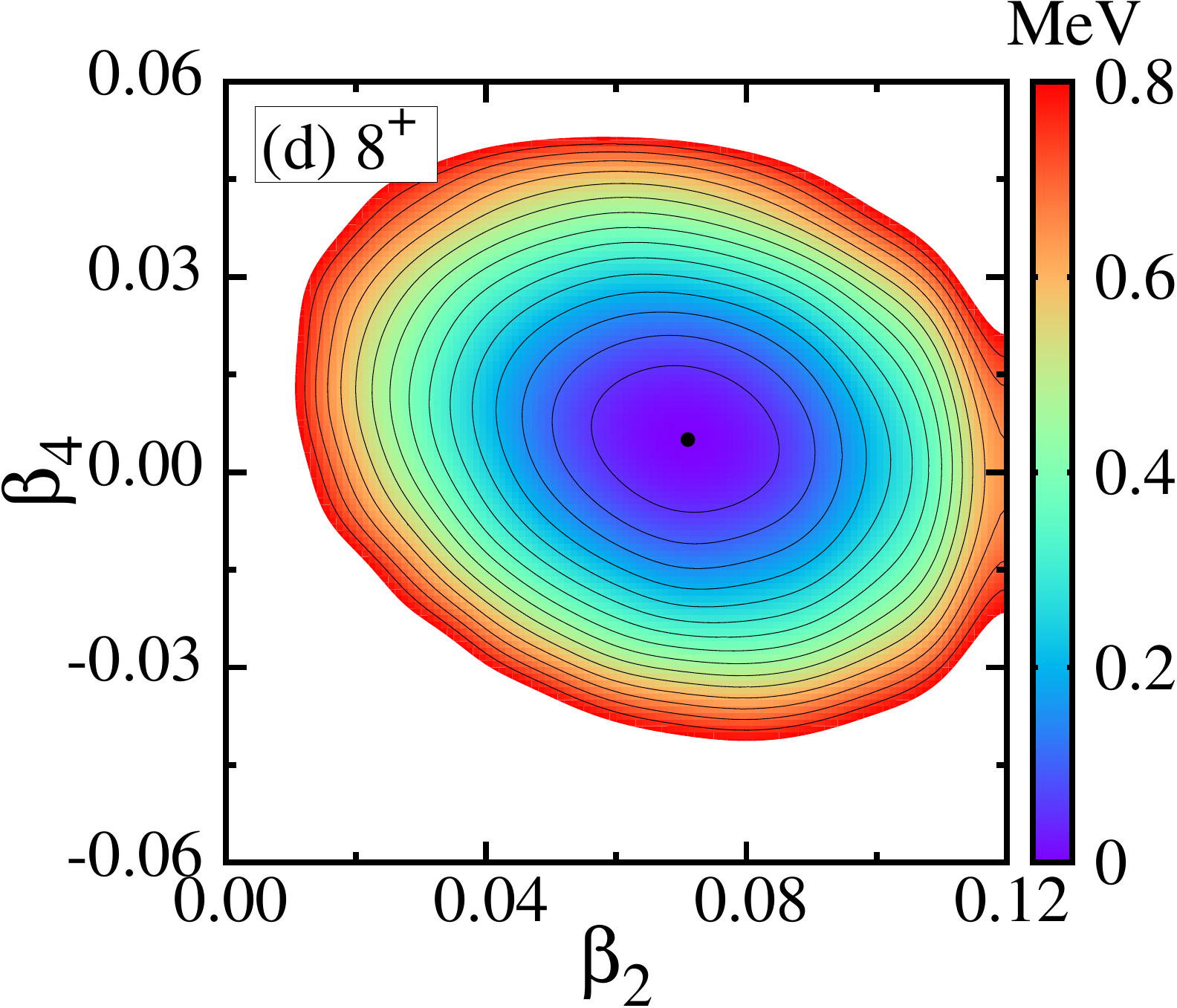}\\
	\caption{Calculated PESs in ($\beta_2$, $\gamma$) and ($\beta_2$, $\beta_4$) planes for ground state (a) and (c), and $K^{\pi}=8^{+} \{\nu7/2^{-}[503]$ $\otimes$ $\nu9/2^{-}[505]\}$ isomeric state (b) and (d) for $^{160}_{76}$Os$_{84}$. According to the Lund convention~\cite{Andersson1976}, the Cartesian coordinates $X = \beta_2$cos($\gamma +30^\circ$) and $Y = \beta_2$sin($\gamma +30^\circ$) are adopted in subplots (a) and (b), avoiding that the minimum appears at the boundary. All the energy intervals between neighboring contour lines are 50 keV. The circle dot denotes the energy minimum which is normalized to zero. More details, see the text, e.g., cf. Table~\ref{table2}.
		                                                          \label{fig4}
	}
\end{figure}
%
Of course, as mentioned in Sec.~\ref{Introduction}, the $\nu 2f_{7/2}$ and $\nu 1h_{9/2}$ orbitals correspond to the $l+1/2$ ($l = 3$) and $l-1/2$ ($l = 5$)  signatures of the spin doublets, respectively. The energy difference between such a pair of orbitals will depend sensitively on the strength of spin-orbit coupling which will lead to not only different changing amplitudes but also different changing directions for them. To evaluate the uncertainty of the spin-orbit coupling, based on the available data in the first and second closest magic nuclei $^{146}$Gd and $^{132}$Sn as seen in Table~\ref{table1}, Fig.~\ref{fig3} illustrates the discrepancies between theoretical and experimental spin-orbit splitting of several spin doublets ($n,l,j=l \pm 1/2$) in functions of spin-orbit strength $\lambda$ for neutrons. The energy difference $\Delta E (n,l)$, indicating the spin-orbit splitting, is defined as $|E(n,l,j=l-1/2)-E(n,l,j=l+1/2)|$ for both experimental and calculated values~\cite{Zalewski2008}. At each $\lambda$ value, the StkI parameters are adopted for other WS parameters during the calculation of theoretical spin-orbit splitting $\Delta E_\texttt{theo.}$. The zero point for each $nl$ orbital indicates that the theoretical $\Delta E_\texttt{theo.}$ equals the experimental $\Delta E_\texttt{exp.}$, corresponding to the optimal spin-orbit strength $\lambda$ for this orbital. From this figure, one can see that, in $^{146}$Gd and $^{132}$Sn, there are five $nl$ orbitals which can be used for extracting the spin-orbit strength $\lambda$ and evaluating its uncertainty. For instance, the optimal $\lambda$ values from $1h$ ($^{146}$Gd), $1h$ ($^{132}$Sn), $2f$ ($^{132}$Sn), $2d$ ($^{132}$Sn) and $3p$ ($^{132}$Sn) orbitals are approximately 29, 29, 21,19 and 14, respectively. Though the reliability of such a single parameter adjustment is somewhat restricted, it still bears some important information on the large uncertainty of spin-orbit coupling parameter $\lambda$. Recently, in Ref.~\cite{Gaamouci2021,Yang2022,Yang2022b} employing phenomenological Woods-Saxon model, a new spin-orbit strength $\lambda = 26.210 (0.513)$ is suggested after removing parametric correlations.       

\subsection{Isomeric structure and its enhanced stability}
\label{pescal}

\begin{table*}
\caption{
Calculated deformations and excitation energies for ground state (g.s.) and two-quasi-neutron ($K^{\pi}=8^{+}$, $\nu{7/2^{-}[503] \otimes 9/2^{-}[505]}$) state in even-even $^{160}\rm{Os}$ and $^{162}\rm{Pt}$, along with part of theoretical and experimental results~\cite{Moller2016,Briscoe2023,Yang2024} for comparison. To evaluate the relative stability, the measured half lives for the ground state and $K^\pi = 8^+$ isometric state are also listed here. Excitation energies ($E^{ex}$) are in MeV.}
\begin{ruledtabular}
	\begin{tabular}{ccccccccccccc}
		\specialrule{0em}{2pt}{2pt}
			\multirow{2}{*}{Nuclei}&\multicolumn{3}{c}{g.s. (this work)}
			&\multicolumn{4}{c}{$K^\pi =8^+$ (this work)}
			&\multicolumn{2}{c}{g.s. (Ref.~\cite{Moller1995})}
			&\multicolumn{3}{c}{Exp.(Ref.~\cite{Briscoe2023})} \\
			\specialrule{0em}{2pt}{2pt} \cline{2-4}  \cline{5-8} \cline{9-10} \cline{11-13}
			\specialrule{0em}{2pt}{2pt}
			&$\beta_{2}$ &$\gamma$  &$\beta_{4}$
			 &$\beta_{2}$ &$\gamma$  &$\beta_{4}$  &$E^{ex}$
             &$\beta_{2}$&$\beta_{4}$
			&$E^{ex}$ &$T_{1/2}$($8^{+}$) & $T_{1/2}$(g.s.)\\
			\specialrule{0em}{2pt}{2pt} \hline
			\specialrule{0em}{2pt}{2pt}
			${}^{160}_{76}\rm{Os}_{84}$  &0.047 &41$^\circ$ &-0.003  &0.071 &60$^\circ$ & 0.005  &1.885
            & 0.064 &-0.010
			&1.844(0.018) &41$^{+15}_{-9} \mu s$ &97$^{+97}_{-32} \mu s$ \\
			\specialrule{0em}{1pt}{1pt}
            ${}^{162}_{78}\rm{Pt}_{84}$  &0.001&56$^\circ$&0.000   &0.062 &60$^\circ$ & 0.005  &2.029 &-&- &- &-&-\\
\specialrule{0em}{2pt}{2pt}
\end{tabular}
\end{ruledtabular}
	                                                             \label{table2}
\end{table*}

%
Figure~\ref{fig4} plots the ground-state PESs and the configuration-constrained ones for two-neutron $K^\pi = 8^+$, $7/2^{-}[503]$ $\otimes$ $9/2^{-}[505]$ isomeric state for $^{160}_{76}$Os$_{84}$, indicating the equilibrium deformations and the softnesses along different deformation degrees of freedom. Let us illustrate that, adopting the Lund convention~\cite{Andersson1976}, the three $\gamma$ sectors [$-120^\circ$, $-60^\circ$], [$-60^\circ$, $0^\circ$] and [$0^\circ$, $60^\circ$] are equivalent and any one is enough to describe the triaxial shape in the static case (below, the calculated $\gamma$ value will be given in the range $0^\circ \leq \gamma \leq 60^\circ$). The ground state possesses a weak oblate deformation, which is enhanced by the polarization effects of these two high-$j$ high-$\Omega$ orbitals in the high-$K$ state. Note that, as illustrated in Ref.~\cite{Xu1998}, during the configuration-dependent PES calculations for a given multi-qp state, the diabatic blocking is performed, namely, when changing deformation, to always follow and block the given orbitals which are occupied by the specified quasiparticles. Table~\ref{table2} lists the calculated results of ground states and high-$K$ states in $^{160}_{76}$Os$_{84}$ and $^{162}_{78}$Pt$_{84}$, along with the available experimental and theoretical data for comparison. For $^{160}_{76}$Os$_{84}$, our calculated excitation energy, 1.885 MeV, is close to the measured value of 1.844 MeV~\cite{Briscoe2023}. The triaxial deformation parameter $\gamma$ is approximately $60^\circ$ at the high-$K$ state, indicating an axially oblate shape. Our calculated $\beta_4$ deformation possesses smaller amplitude than that given in Ref.~\cite{Moller1995} but has the same signs. The combination of high $K$, low energy, and axially deformed shape provides the necessary conditions for the formation of isomers.  
Indeed, in the nucleus $^{160}$Os, the $K^\pi = 8^{+}$ isomer has been observed to possess a half-life of $40^{+15}_{-9}\mu s$, which is of the same order of magnitude as its ground-state half-life $97^{+97}_{-32}\mu s$~\cite{Briscoe2023}. It should be mentioned that the calculated $\beta_2$ values are soft at ground states in these two nuclei but stabilized at oblate two-neutron high-$K$ states, e.g., cf. Fig.~\ref{fig4}.

\begin{figure*}[htbp]
	\centering
	\includegraphics[height=0.46\textwidth]{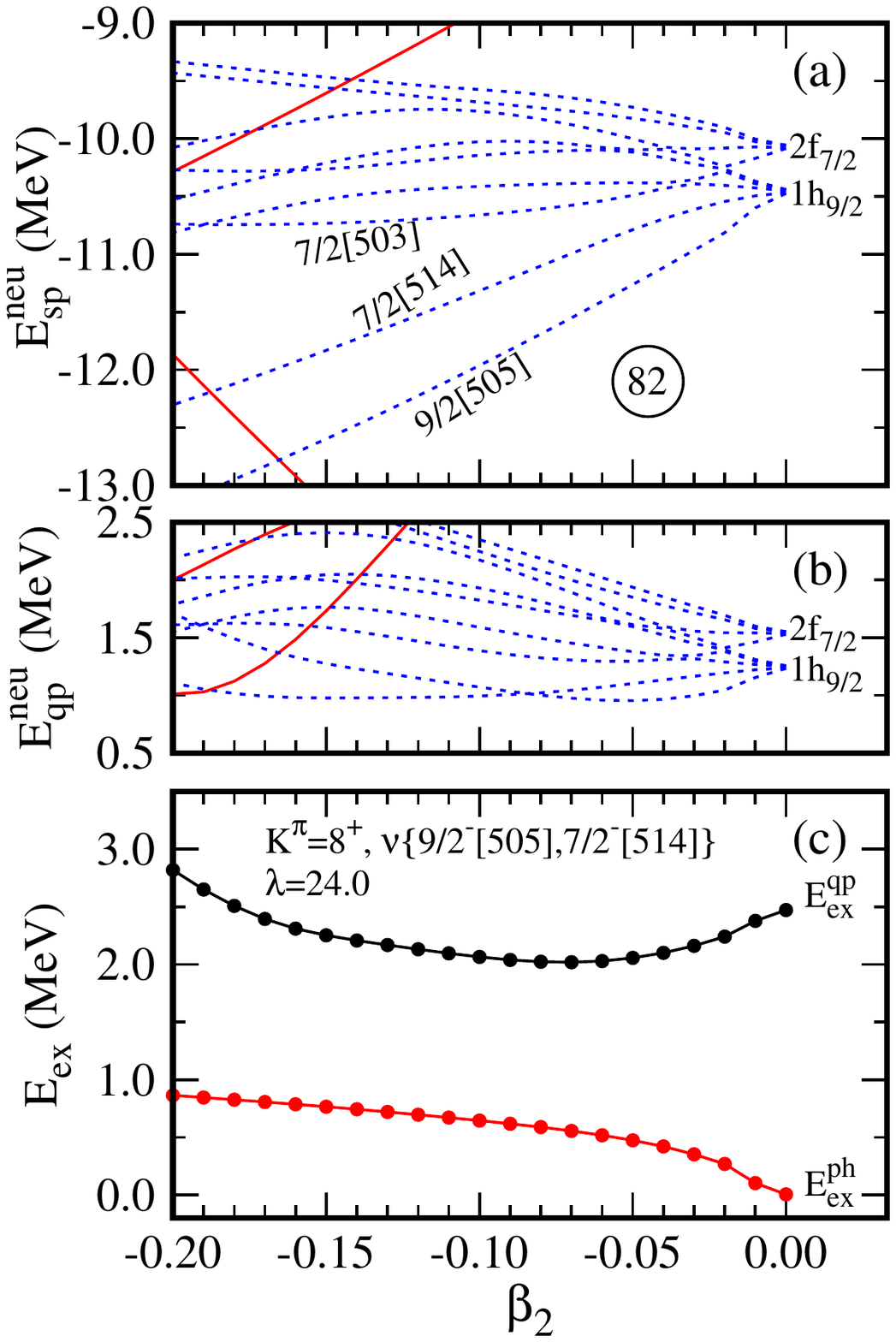}
	\includegraphics[height=0.46\textwidth]{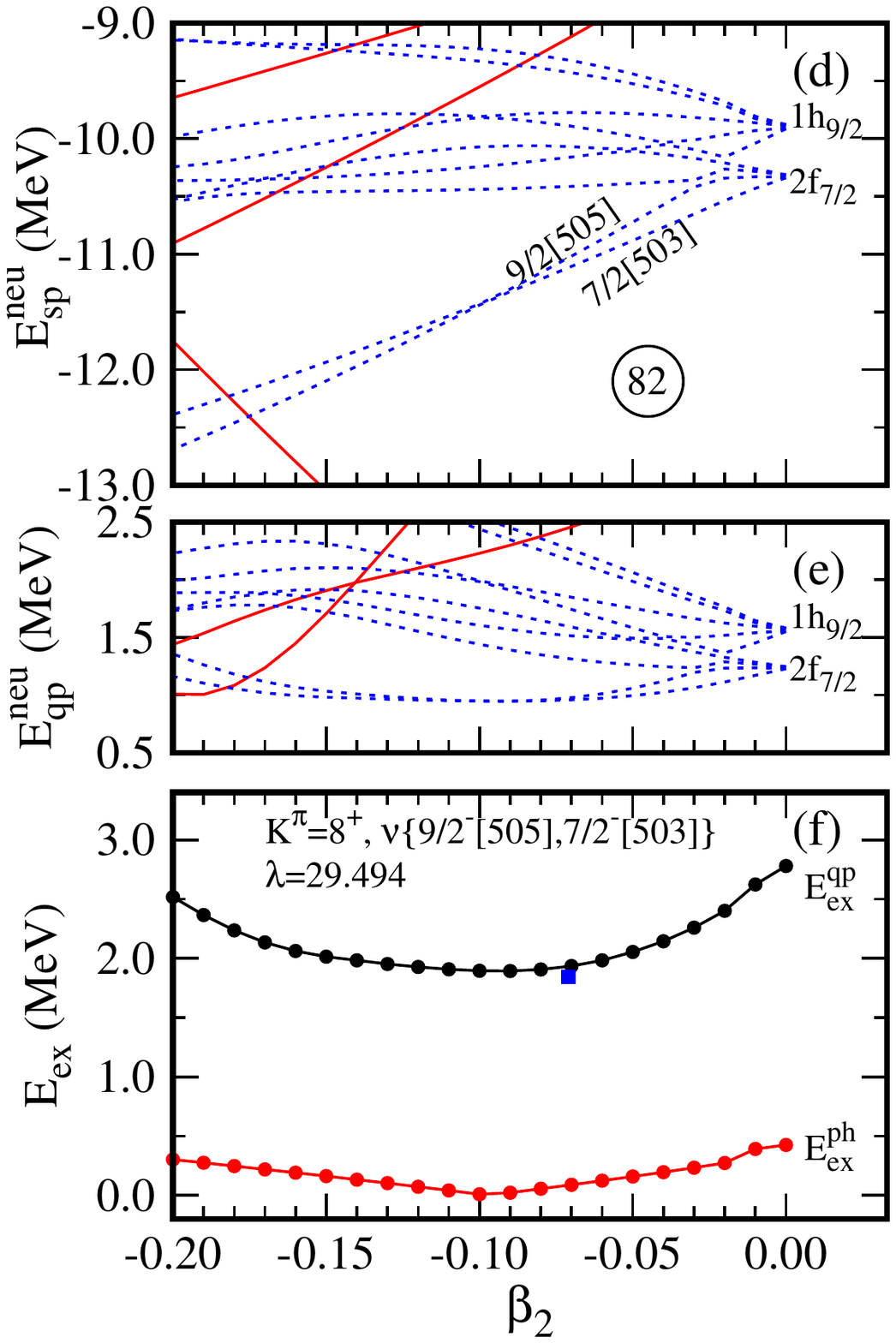}
	\includegraphics[height=0.46\textwidth]{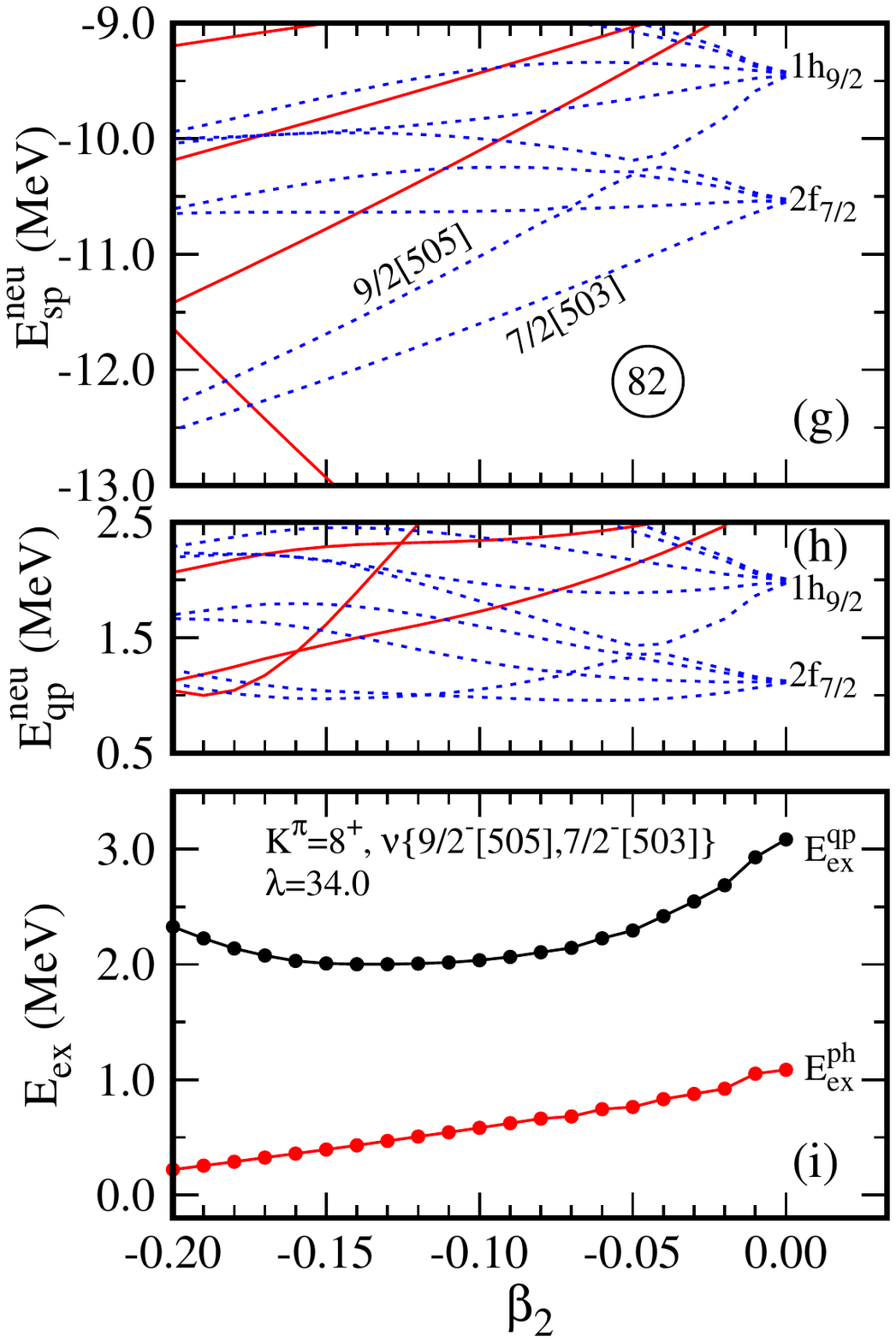}
	\caption{
	Calculated single-particle levels (a), (d), (g), quasi-particle levels (b), (e), (h) and excitation energies (c), (f), (i) of $K^\pi = 8^+$ isomer in functions of $\beta_2$ at three selected strengths of spin-orbit coupling (namely, $\lambda$ = 24.0, 29.494 and 34.0) for $^{160}$Os. Note that the single- and quasi-particle levels with red solid and blue dotted lines correspond to positive- and negative-parity states, respectively. The negative $\beta_2$ deformation indicates the oblate shape, realizing by fixing $\gamma$ to $60^\circ$ in the calculations. The blue squared symbol at $\beta_2=0.071$ in (f) denotes the observed excitation energy of $K^\pi = 8^+$ isomeric state. See the context for more explanations.
		                                                           \label{fig5}
	}
\end{figure*}

Oblate high-$K$ isomers are relatively rare and of interest. The $N= 84$ isotones, whose neutron Fermi surfaces lie above the  $N = 82$ shell closure and between two high-$\Omega$ orbitals, e.g., $7/2^{-}[503]$ and $9/2^{-}[505]$, at the oblate side, provide us a precious playground to investigate the high-$K$ low-energy isomers. Nevertheless, as mentioned above, such high-$\Omega$ orbitals primarily originate from the different spin doublets with different angular momentum numbers [namely, ($l+1/2$; $l=3$) and ($l-1/2$; $l=5$) signatures].
The spin-orbit interaction will strongly affect their relative energies, further changing the excitation energy and wave-function structure of the high-$K$ isomeric state.

\begin{table*}
\caption{
The interested neutron single-particle levels near the Fermi surface calculated at the selected weakly-oblate deformation point $\beta_{2}= -0.05$ (equivalently, at $\beta_2= 0.05$ and $\gamma= 60.00^\circ$ in the Bohr parameterization) in the nucleus $^{160}_{76}$Os$_{ 84}$, along with the first three wave-function (w.f.) components expanded on the cylindrical basis $|Nn_{z}\Lambda\Omega\rangle$ (upper)  and spherical basis $|Nlj\Omega\rangle$. The calculations are performed by using the Woods-Saxon Hamiltonian with the StkI parameters~\cite{Meng2018,Bhagwat2010,Bhagwat2023} but different spin-orbit strengths ($\lambda=24.0$, 29.494 and 34.0). Let us remind that, for comparison, a set of quantum numbers \{$Nlj\Omega$\} in the Dirac notation is adopted here, different from \{$nlj$\} as seen, e.g., in Figs.1-3 and 5 (but equivalent due to $N=2(n-1)+l$).	\\ }
\begin{ruledtabular}
\begin{tabular}{lclc}
\multirow{1}{*} {$\lambda $} & $e$(MeV)  & Label & First three w.f. components \\ 
\specialrule{0.04em}{3pt}{3pt}\\
\multirow{1}{*}{$24.0$}
            & -11.26 & $| 5 0 5\frac{ 9}{2}\rangle$ & 87.8\% $| 5 0 5\frac{ 9}{2}\rangle$+11.9\% $| 5 1 4\frac{ 9}{2}\rangle$+ 0.2\% $|11 0 5\frac{ 9}{2}\rangle$ \\
\specialrule{0.00em}{3pt}{3pt}
             &       &$| 5h_{ 9/2}\frac{ 9}{2}\rangle$ & 99.4\% $| 5h_{ 9/2}\frac{ 9}{2}\rangle$+ 0.2\% $|11h_{ 9/2}\frac{ 9}{2}\rangle$+ 0.2\% $| 5h_{11/2}\frac{ 9}{2}\rangle$ \\
\specialrule{0.00em}{3pt}{6pt}
             &-10.79 &$| 5 1 4\frac{ 7}{2}\rangle$ & 50.6\% $| 5 1 4\frac{ 7}{2}\rangle$+44.4\% $| 5 0 3\frac{ 7}{2}\rangle$+ 2.1\% $| 3 0 3\frac{ 7}{2}\rangle$ \\
             \specialrule{0.00em}{3pt}{3pt}
             && $| 5h_{ 9/2}\frac{ 7}{2}\rangle$ & 56.9\% $| 5h_{ 9/2}\frac{ 7}{2}\rangle$+38.1\% $| 5f_{ 7/2}\frac{ 7}{2}\rangle$+ 2.7\% $| 3f_{ 7/2}\frac{ 7}{2}\rangle$ \\
             \specialrule{0.00em}{3pt}{6pt}
             & -10.49 & $| 5 0 3\frac{ 7}{2}\rangle$ &35.3\% $| 5 0 3\frac{ 7}{2}\rangle$+31.2\% $| 5 2 3\frac{ 7}{2}\rangle$+27.8\% $| 5 1 4\frac{ 7}{2}\rangle$ \\
             \specialrule{0.00em}{3pt}{3pt}
             &&$| 5f_{ 7/2}\frac{ 7}{2}\rangle$ &51.0\% $| 5f_{ 7/2}\frac{ 7}{2}\rangle$+42.5\% $| 5h_{ 9/2}\frac{ 7}{2}\rangle$+ 3.0\% $| 3f_{ 7/2}\frac{ 7}{2}\rangle$ \\
 \specialrule{0.00em}{3pt}{3pt} \cline{1-4}
\specialrule{0.00em}{3pt}{3pt}
\multirow{1}{*}{$29.494$}
            &-10.88 &$| 5 0 3\frac{ 7}{2}\rangle$ &79.3\% $| 5 0 3\frac{ 7}{2}\rangle$+ 7.8\% $| 5 2 3\frac{ 7}{2}\rangle$+ 5.5\% $| 3 0 3\frac{ 7}{2}\rangle$ \\
             \specialrule{0.00em}{3pt}{3pt}
            &&$| 5f_{ 7/2}\frac{ 7}{2}\rangle$ & 85.9\% $| 5f_{ 7/2}\frac{ 7}{2}\rangle$+ 5.9\% $| 3f_{ 7/2}\frac{ 7}{2}\rangle$+ 4.3\% $| 7f_{ 7/2}\frac{ 7}{2}\rangle$ \\
            \specialrule{0.00em}{3pt}{6pt}
            & -10.72 &$| 5 0 5\frac{ 9}{2}\rangle$ & 88.3\% $| 5 0 5\frac{ 9}{2}\rangle$ +11.4\% $| 5 1 4\frac{ 9}{2}\rangle$+ 0.2\% $|11 0 5\frac{ 9}{2}\rangle$ \\
            \specialrule{0.00em}{3pt}{3pt}
            &&$| 5h_{ 9/2}\frac{ 9}{2}\rangle$ & 99.5\% $| 5h_{ 9/2}\frac{ 9}{2}\rangle$ + 0.2\% $|11h_{ 9/2}\frac{ 9}{2}\rangle$+ 0.1\% $| 5h_{11/2}\frac{ 9}{2}\rangle$ \\
            %
 \specialrule{0.00em}{3pt}{3pt} \cline{1-4}
  \specialrule{0.00em}{3pt}{3pt}
 \multirow{1}{*}{$34.0$}
            & -11.07&$| 5 0 3\frac{ 7}{2}\rangle$ &78.6\% $| 5 0 3\frac{ 7}{2}\rangle$+10.0\% $| 5 2 3\frac{ 7}{2}\rangle$+ 5.9\% $| 3 0 3\frac{ 7}{2}\rangle$ \\
           \specialrule{0.00em}{3pt}{3pt}
            &&$| 5f_{ 7/2}\frac{ 7}{2}\rangle$ &87.3\% $| 5f_{ 7/2}\frac{ 7}{2}\rangle$+ 6.2\% $| 3f_{ 7/2}\frac{ 7}{2}\rangle$+ 4.6\% $| 7f_{ 7/2}\frac{ 7}{2}\rangle$ \\
           \specialrule{0.00em}{3pt}{6pt}
           & -10.29 & $| 5 0 5\frac{ 9}{2}\rangle$  & 88.5\% $| 5 0 5\frac{ 9}{2}\rangle$ + 11.1\% $| 5 1 4\frac{ 9}{2}\rangle$+ 0.2\% $|11 0 5\frac{ 9}{2}\rangle$ \\
           \specialrule{0.00em}{3pt}{3pt}
           & &$| 5h_{ 9/2}\frac{ 9}{2}\rangle$ & 99.5\% $| 5h_{ 9/2}\frac{ 9}{2}\rangle$+ 0.2\% $|11h_{ 9/2}\frac{ 9}{2}\rangle$+ 0.1\% $| 9h_{ 9/2}\frac{ 9}{2}\rangle$ \\
			\specialrule{0.00em}{3pt}{6pt}
		\end{tabular}
	\end{ruledtabular}
	                                                               \label{table3}
\end{table*}

Taking the uncertainty of spin-orbit coupling shown in Fig.~\ref{fig3} into account, Figure~\ref{fig5} respectively shows, focusing on the oblate side, the single-particle levels, the quasi-particle levels and the excitation energies (including particle-hole and quasiparticle excitations) of $K^\pi = 8^+$ isomer calculated at three selected strengths of spin-orbit coupling, namely, $\lambda = 24.0, 29.494$ (the original value) and 34.0. Note that the single-particle levels at each $\beta_2$ are calculated by the deformed WS potential with the StkI parameters, except for changing $\lambda$. The quasi-particle levels are obtained by considering the pairing effects. The excitation energies based on the quasi-particle and particle-hole schemes for $K^\pi = 8^+$ isomer are respectively obtained by calculating the summation of the excited quasi-particle levels and the difference between the excited particle and hole levels, ignoring the deformation effect before and after the excitation~\cite{Walker2024}. Such an approximate treatment can help to intuitively understand and evaluate the effect of spin-orbit coupling on the structural properties in this high-$K$ isomeric state. For instance, the first high-$\Omega$ orbitals above and below the $N = 84$ neutron Fermi surface will vary at the low $\lambda$, e.g., see Fig.~\ref{fig5}(a), indicating a possible configuration change of the low-lying $K^\pi =8^+$ isomer. With the increase of the spin-orbit strength, the levels near the Fermi surface exhibit a different and interesting evolution. As seen in Figs.~\ref{fig5}(a), ~\ref{fig5}(d) and ~\ref{fig5}(g), two high-$\Omega$ orbitals will respectively approach, cross and depart each other with the increasing $\beta_2$, which will further lead the one-particle one-hole (1p-1h) excitation energy $E^{ph}_{ex}$ to the evolution trend in Fig.~\ref{fig5}(c), \ref{fig5}(f) and \ref{fig5}(i), respectively. That is, the energy $E^{ph}_{ex}$ always decreases (increases) as $\beta_2$ increases from -0.20 to 0.00 in Fig.~\ref{fig5}(c) [Fig.~\ref{fig5}(i)], but in Fig.~\ref{fig5}(f) it decreases firstly and then increases. Needless to say, without the inclusion of the pairing effect, the 1p-1h excitation energy $E^{ph}_{ex}$ is somewhat smaller than the experimental data. Considering the quasi-particle scheme with the pairing effect, the low-lying two-quasi-particle excitation energy $E^{qp}_{ex}$ of $K^\pi = 8^+$ isomeric state is also illustrated at different $\lambda$, as seen in Fig.~\ref{fig5}(c), \ref{fig5}(f) and \ref{fig5}(i), in good agreement with data to a large extent, e.g., within the error bar of 0.2 MeV. In fact, adopting the universal WS parameters, to change the spin-orbit coupling strength may to a large extent reproduce the similar evolution properties for this $K^\pi = 8^+$ isomer.

Table~\ref{table3} presents the calculated wave-function components of the interested levels $7/2^{-}[503]$ and $9/2^{-}[505]$ (including $7/2^{-}[514]$ at $\lambda = 24.0$) at three selected spin-orbit strengths near the equilibrium deformations in $^{160}$Os. The single-particle label is denoted by the largest component of the corresponding basis function, e.g., the harmonic oscillator (h.o.) basis in the cylindrical and spherical coordinate systems. It should be noticed that the largest components in these two high-$j$ high-$\Omega$ orbitals $7/2^{-}[503]$ (or $|5f_{7/2}\frac{7}{2}>$) and $9/2^{-}[505]$ (or $|5h_{9/2}\frac{9}{2}>$) are dominated in the cases of $\lambda = 29.494$ and 34.0. As one can see that their percentages are all more than 78\%, even 99\% for $|5h_{9/2}\frac{9}{2}>$.    However, at $\lambda = 24.0$, the mixing of the basis-function components $7/2^{-}[503]$ and $7/2^{-}[514]$ is serious for the $-10.79$- and $-10.49$-MeV single-particle levels, with a small percentage difference between the first two components both in the cylindrical and spherical expansions of the h.o. basis functions. In this circumstance, an arbitrary configuration assignment for the two-quasi-particle $K^\pi = 8^+$ isomeric state will be somewhat risky, in particular, when considering the virtual level crossing due to the level non-crossing rule~\cite{Hatton1976,Bengtsson1988}.          

\begin{figure}[htbp]
	\centering
	\includegraphics[height=0.35\textwidth]{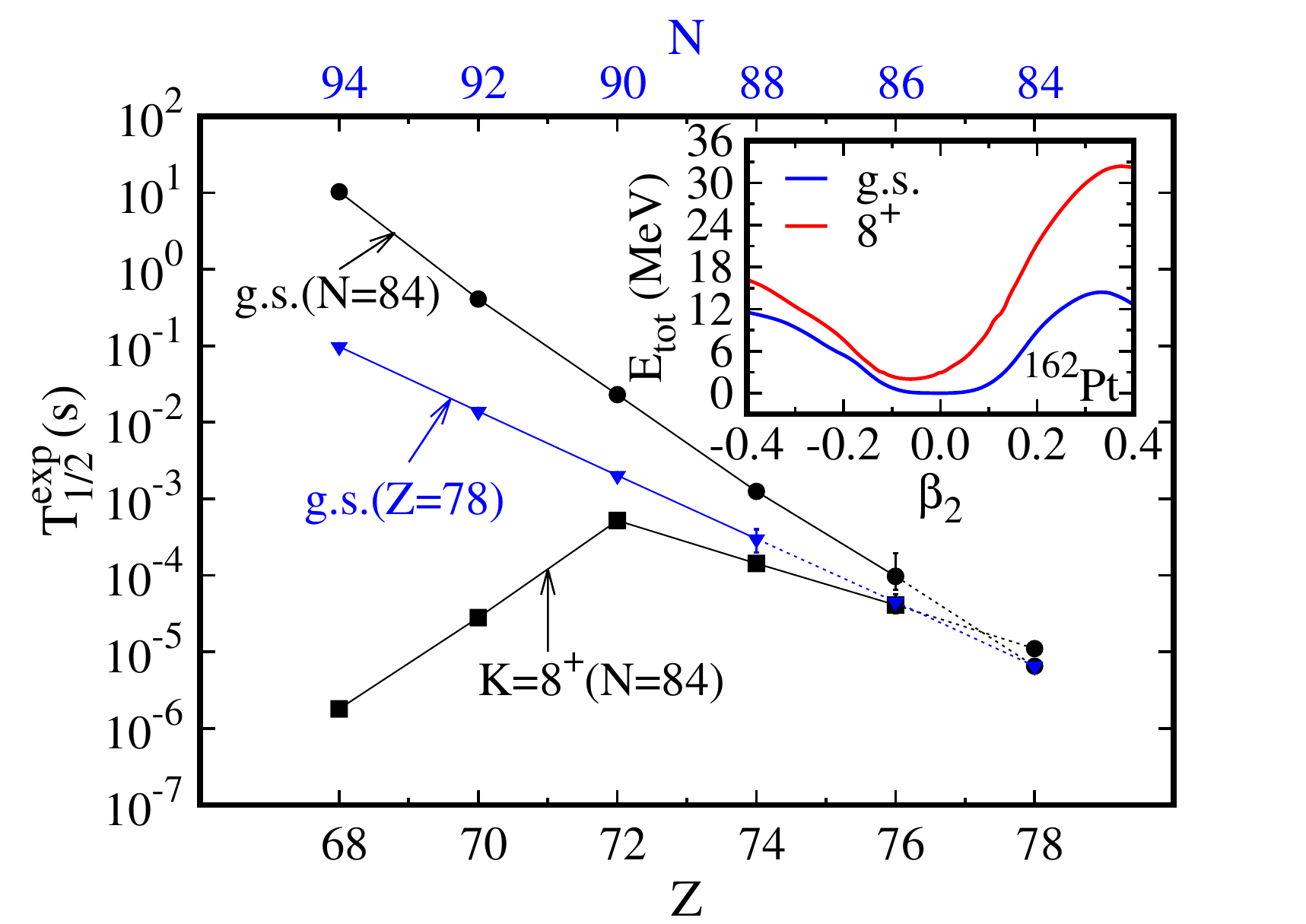}
	\caption{
Experimental half-lives of the ground states (circles) and two-quasiparticle $K^\pi = 8^+$ isomeric states (squares) for the $N=84$ isotones with $68 \leq Z \leq 76$, along with the ground-state values for the $Z = 78$ isotopes. The half lives, connected by the dotted lines, e.g., for the nucleus $^{162}_{78}$Pt${_{84}}$ (undiscovered) are obtained by simple linear extrapolation.  Note that, except for the ground-state half lives in $^{160}$Os and $^{166}$Pt, the error bars are too small to be visible. Calculated potential energies in functions of $\beta_2$ for the ground state and the $K^\pi = 8 ^+$ isomer in $^{162}_{78}$Pt${_{84}}$ are illustrated in the inset. For more details, see the text.
		\label{fig6}
	}
\end{figure}


In the $N=84$ isotones, the lifetimes of the isomeric states, comparing with those of the ground states, have a relatively enhanced trend with increasing proton number $Z$ though both of them decrease~\cite{Clark2019,NNDC2025}. Figure 6 shows the half lives of the ground states and the observed $K^\pi = 8^+$ isomeric states in the $N=84$ even-even isotones ranged from $Z = 68$ to $78$, together with the ground-state half-lives of the $Z = 78$ even-even isotopes with $88 \leq N \leq 94$. From Fig.~\ref{fig6}, one can see that the ground-state half-lives in both the isotones and isotopes exhibit good linear relationships and the extrapolated lifetimes from the isotonic and isotopic directions for $^{162}_{78}$Pt$_{84}$ are in good agreement with each other. Let us remind that, as observing the half-life evolution of the $Z = 78$ isotopes, one should use the top horizontal coordinate $N$. Though the most neutron-deficient nucleus that has been so far discovered is $^{165}_{78}$Pt~\cite{NNDC2025}, according to HFB-14 mass model, which is based on the Hartree-Fock-Bogoliubov method with Skyrme forces and a $\delta$-function pairing force~\cite{Goriely2007}, three more isotopes $^{162-164}$Pt are still predicted to be particle-stable along the proton dripline~\cite{Amos2011}. That is to say, the nucleus $^{162}_{78}$Pt$_{84}$ may be accessible in experiment, with a possible half-life above the microsecond order of magnitude, as seen in Fig.~\ref{fig6}. According to the evolution trends of the half lives for ground states and $K^\pi = 8^+$ isomeric states, the stability between ground state and isomeric state inverses in $^{162}_{78}$Pt$_{84}$. 

To understand the enhanced stability of the high-$K$ isomer, as seen in the inset of Fig.~\ref{fig6}, we show the calculated potential-energy curves of the ground and $K^\pi = 8^+$ isomeric states for the nucleus $^{162}_{78}$Pt$_{84}$. Both two curves are normalized to the ground-state minimum and the calculations at $-\beta_2$ points are performed by the energy minimization near $\gamma=60^\circ$. In this nuclear region, $\alpha$-decay is the primary decay mode~\cite{NNDC2025} and the $\alpha$-decay width is proportional to the preformation probability of an $\alpha$ particle at the nuclear surface and the $\alpha$ penetration probability through the potential barrier~\cite{Gurney1928}. Indeed, for the two-quasi-particle high-$K$ configuration, the $\alpha$ preformation factor may have a considerable reduction due to the decreased pairing density~\cite{Xu2004,Poggenburk1969}. In addition, the increased barrier height and width, e.g., see the inset of Fig.~\ref{fig6}, also prefer reducing the $\alpha$-emission probability and increasing the half life. The detailed survey of the corresponding lifetimes may be of interest but beyond the scope of the present work. However, similar to the superheavy mass region~\cite{Xu2004}, one can still conclude from the facts mentioned above that the high-$K$ isomer may enhance the survival probabilities of drip-line nuclei with sufficiently short lifetimes and help to extend the nuclear chart. It seems that $^{162}_{78}$Pt$_{84}$ could be the possible candidate to probe the lifetime inversion between ground and isomeric states in the proton drip-line region.     

\section{Summary}
\label{summary}
The recently experimental extension in $^{160}\rm{Os}$ provides us an opportunity to further study nuclear structure in the most neutron-deficient $N = 84$ isotone up to now. In this project, both ground-state and $K^{\pi}=8^{+}$ isomeric-state properties in the two-proton drip-line nucleus $^{160}\rm{Os}$ have been investigated by using conﬁguration constrained PES calculations in multidimensional ($\beta_2, \gamma, \beta_4$) deformation space. 
The uncertainty of the spin-orbit coupling strength is evaluated and its impact on two-quasiparticle $K^{\pi}=8^{+}$ isomeric structure is estimated. 
The energy crossing or inversion of high-$\Omega$ orbitals ($7/2^-[503]$ and $9/2^-[505]$ orbitals) is analyzed by comparing the microscopic single-particle structures with different spin-orbit coupling parameters. The main wave-function components for the interested single-particle levels, e.g., $f_{7/2}\frac{7}{2}$ and $h_{9/2}\frac{9}{2}$, are illustrated to exhibit configuration structure and its evolution with spin-orbit strength. The excitation energies of the $K^{\pi}=8^{+}$ isomer based on quasiparticle and particle-hole excitations are provided and compared with data, revealing the critical roles of the pairing interaction. 
Both experimentally and theoretically, it is to an extent meaningful to further investigate the multi-quasiparticle isomers, which may have the enhanced stability in the drip-line nuclei. Along the $N=84$ isotonic chain, the next neutron-deficient even-even nucleus $^{162}$Pt is suggested as the favorable candidate for verifying the stability inversion between its isomeric and ground states in this proton drip-line region.

\section*{Acknowledgement}

We are deeply grateful to Professor Ramon Wyss for the illuminating discussion, especially on the WS mean field and model parameters. This work was supported by the Natural Science Foundation of Henan Province (No. 252300421478) and the National Natural Science Foundation of China (No.11975209, No. U2032211, and No. 12075287). Some of the calculations were conducted at the National Supercomputing Center in Zhengzhou.

The authors declare that they have no known competing financial
interests or personal relationships that could have appeared to
influence the work reported in this paper.

\section*{References}
%

\end{document}